\documentstyle[12pt]{article}
\begin{document}

\begin{center}
{\large\bf THE FRENET SERRET DESCRIPTION OF GYROSCOPIC PRECESSION}\\~\\
B.R.Iyer\footnote{e-mail: bri@rri.ernet.in}\\
Raman Research Institute, \\
Bangalore 560080, India\\~\\
C.V.Vishveshwara\footnote{e-mail: vishu@iiap.ernet.in}\\
Indian Institute of Astrophysics, \\
Bangalore 560034, India
\end{center}
\begin{abstract}
\noindent
The phenomenon of gyroscopic precession is studied within the framework of
Frenet-Serret formalism adapted to quasi-Killing trajectories. Its
relation to the congruence vorticity is highlighted with particular
reference to the irrotational congruence admitted by the stationary,
axisymmetric spacetime. General precession formulae are obtained for
circular orbits with arbitrary constant angular speeds. By successive
reduction, different types of precessions are derived for the Kerr -
Schwarzschild - Minkowski spacetime family. The phenomenon is studied in
the case of other interesting spacetimes, such as the De Sitter and G\"{o}del
universes as well as the general stationary, cylindrical, vacuum
spacetimes.
\end{abstract}
\vspace{2cm}
{\em To Appear in Phys. Rev. D 1993}
\newpage
\section{Introduction}
The phenomenon of rotation exhibits interesting and often intriguing
physical effects. This is even more so within the framework of the general
theory of relativity which leads to novel features. These features, for
instance, are built into the structure of spacetime, such as that of a
rotating black hole. Dragging of inertial frames is a typical example of
rotational effects incorporated into the spacetime structure. Such
effects also manifest themselves in the intrinsic aspects of particle
motion and related phenomena like the gyroscope precession. These aspects
can be elegantly studied by the invariant geometrical description of
particle trajectories that follow the directions of spacetime symmetries,
or Killing vector fields, provided of course that the spacetime admits
such symmetries. This is accomplished by adopting the Frenet-Serret
formalism to characterize the Killing trajectories of a four
dimensional spacetime. Of the three geometric parameters basic to this
formalism, the curvature is identified with the particle acceleration,
while the two torsions are directly related to the gyroscope precession.
Furthermore, the Frenet-Serret tetrad provides a convenient reference
frame for the description of all relevant physical phenomena. Therefore,
when the formalism is applied to the timelike integral curves of
 spacetime symmetries, the phenomenon of gyroscope precession can be completely
analysed in a natural and cogent manner.

The Frenet-Serret formalism applied to the Killing trajectories can be
extended in a straightforward manner to what we may term as a quasi-Killing
congruence. This congruence consists of
timelike curves following the direction given by a combination of Killing
vectors with non-constant coefficients. An important example is the
irrotational congruence admitted by the Kerr spacetime. With the help of
this extended formalism, a broad based framework is provided for the
study of gyroscope precession in a variety of circumstances.

The present paper is organised as follows. In section 2, we discuss the
application of Frenet-Serret formalism to the quasi-Killing trajectories,
precession of gyroscopes transported along them and its relation to the
vorticity of the congruence. Section 3 considers the stationary, axially
symmetric spacetimes and
concentrates on the globally timelike Killing trajectories followed by
stationary observers. Specializing to the Kerr spacetime, the gyroscopic
precession with respect to the stationary observer - a direct
manifestation of inertial frame dragging - is displayed. By using rotating
coordinates, gyroscopic precession along circular orbits with arbitrary
constant angular
speeds is investigated in section 4. The general formulas derived are
first applied to the Kerr spacetime to obtain particle acceleration and
gyroscopic precession without approximation. Then by successive
specialization, we obtain the Schiff  precession, precession in the
Schwarzschild spacetime with Fokker-De Sitter precession as a particular
example, and Thomas precession in the Minkowski spacetime. The irrotational
congruence is discussed and the Frenet-Serret parameters are derived for
the corresponding trajectories. The general formalism is also applied to
the De Sitter spacetime. Section 5, treats in detail the general case of
stationary cylindrically  symmetric spacetimes where the general
quasi-Killing trajectories are helical orbits. Included in this
study as special cases are the G\"{o}del universe and the general vacuum
metrics
as given by Vishveshwara and Winicour. Section 6 comprises a summary and
concluding remarks.

Starting from Thomas precession, gyroscopic precession has been studied
extensively by different approaches both in special and general
relativity \cite{mtw,tpm,rp90,tes87,ab90,wj92}. We have presented here a
unified, covariant, geometric
treatment of this remarkable phenomenon. Furthermore, in this treatment
computations can  be made in a straightforward and complete
manner. We have also highlighted the inter-relations among quantities such
as vorticity, precession and Frenet-Serret torsions. In addition to the
general formulae, exact expressions have been presented pertaining to
special and physically significant spacetimes. It is hoped that the
discussions and formalism of this paper offer additional insight into the
phenomenon of gyroscopic precession and that the formulas derived can be
of use for further elucidation and astrophysical applications.

Our metric signature is $(+,\, -,\, -,\, -)$. Spacetime indices are denoted by
latin letters $a,\,b\,\ldots m,\,n\,\ldots$. and run over $0,\,1,\,2,\,3 $
while spatial indices are
denoted by greek letters $\alpha, \beta \cdots \mu, \nu \cdots$ and run over
$1,\,2,\,3$. The corresponding tetrad (triad) indices are indicated by
enclosing
them in parentheses $(a) (b) \cdots (m) (n) \cdots ((\alpha) (\beta) \cdots
(\mu) (\nu) \cdots$).

\section{The Quasi-Killing Trajectories}
\subsection{The Frenet-Serret Formalism}
In \cite{hsv74} and \cite{iv88} it was shown that the Frenet-Serret  formalism
has some
attractive formal properties in the case of Killing trajectories that find
elegant applications in black hole geometries. We show
below how these properties obtain in a more general case, we call
quasi-Killing.  Consider a spacetime with a timelike Killing vector
{\boldmath $\xi$} and a
set of spacelike Killing vectors {\boldmath $\eta_{(A)}$} $\,(A = 1,2\cdots
m)$. The
combination
\begin{equation}
\chi^a\,\equiv\,\xi^a\,+\,\omega_{(A)}\eta^{a}_{(A)}
\end{equation}
where summation over $(A)$ is implied and
\begin{equation}
\raisebox{-.25cm}{$\stackrel{\textstyle \cal L}{\scriptstyle
\chi}$}\;\omega_{(A)}\,=\,0
\end{equation}
is called a quasi-Killing vector. The terminology is justified for our
usage since as in the Killing case it follows that, if, $u^a$ is the
four-velocity associated with $\chi^a$ (where it is timelike) obtained by
normalizing $\chi^a$,

\begin{eqnarray}
e^{a}_{(0)}\,&\equiv&\,u^a\,\equiv\,e^\psi\,\chi^a\\
{\mbox{\rm then}}\;\;e^{
-2\psi}\,&=&\,\chi_a\chi^a\;\;;\;\;\psi_{,a}\,\chi^a\,=\,0  \\
{\mbox{{\rm and }}}\;\dot {e}^{a}_{(0)}\,&\equiv &\,e^{a}_{(0);b}\,
e^{b}_{(0)}\,=\,F^{a}_{\;\; b}\,e^{b}_{(0)}\label{eq:1}\\
\mbox{{\rm where}}\;F_{ab}\,&\equiv&\,e^\psi\,(\xi_{a;b}\,+\,\omega_{(A)}\,
\eta_{(A)a;b})\label{eq:2}
\end{eqnarray}

It is easy to show, using the Killing equation and the
relation $\xi_{a;b;c}\,=\,R_{abcd}\,\xi^d$ for any Killing vector
{\boldmath $\xi $},
that:
\begin{equation}
F_{ab}\,=\,-F_{ba}\;\;; \;\;\;\;
\dot{F}_{ab}\,=\,0 \label{eq:3}
\end{equation}
Recall that the Frenet-Serret equations are \cite{hsv74,iv88}
\begin{eqnarray}
\dot{e}^{a}_{(0)}\,&=&\,\kappa\,e^{a}_{(1)}\nonumber\\
\dot{e}^{a}_{(1)}\,&=&\,\kappa\,e^{a}_{(0)}\,+\,\tau_1\,e^{a}_{(2)}\nonumber\\
\dot{e}^{a}_{(2)}\,&=&\,-\tau_1\,e^{a}_{(1)}\,+\,\tau_2\,e^{a}_{(3)}\nonumber\\
\dot{e}^{a}_{(3)}\,&=&\,-\tau_2\,e^{a}_{(2)}\label{eq:4}
\end{eqnarray}
where $\kappa$ is the curvature and $\tau_1,\,\tau_2$ the first and second
torsions respectively. The Frenet-Serret equations (\ref{eq:4}) together with
equations (\ref{eq:1}-\ref{eq:3}) imply, as in the Killing case, that along
trajectories
of $\chi^a$ the Frenet-Serret invariants $\kappa$, $\tau_1$ and  $\tau_2$
are constants
and the Frenet-Serret basis vectors $e^{a}_{(i)}$ satisfy a
Lorentz-like equation:
\begin{eqnarray}
\dot{\kappa}\,&=&\,\dot{\tau}_1\,=\,\dot{\tau}_2\,=\,0\\
\dot{e}^{a}_{(i)}\,&=&\,F^{a}_{\;\;b}\,e^{b}_{(i)}
\end{eqnarray}
Note that $F_{ab}\,\neq\,e^\psi\,\chi_{a;b}$.\\
As before \cite{hsv74,iv88}\\
\begin{eqnarray}
\kappa^2\,&=&\,F^{2}_{ab}\,e^{a}_{(0)}\,e^{b}_{(0)}\label{eq:5}\\
\tau^{2}_{1}\,&=&\,\kappa^2\,-\,\frac{F^{4}_{ab}\,e^{a}_{(0)}\,e_{(0)}^{b}}
{\kappa^2}\label{eq:6}\\
\tau^{2}_{2}\,&=&\,-\,\frac{(\kappa^2-\tau^{2}_{1})^2}{\tau^{2}_{1}}\,+\,
\frac{F^{6}_{ab}\,e^{a}_{(0)}\,e^{b}_{(0)}}{\kappa^2\,\tau^{2}_{1}}\label{eq:7}
\\
{\rm where}\;\; (F^n)_{ab}\,&\equiv&\,F_a^{\;\;a_1}\,F_{a_1}^{\;\;a_2}\cdots
F_{a_{n-1}b}\\
\mbox{{\rm
Moreover,}}\;\;\alpha\,&\equiv&\,\frac{1}{2}\,F^{a}_{\;\;b}\,F^{b}_{\;\;a}\;\;
=\;\;  \kappa^2\,-\,\tau^{2}_{1}\,-\,\tau^{2}_{2}\label{eq:fsi4}
\end{eqnarray}

Before proceeding further we may mention some examples of quasi-Killing
congruences given by (1). In the stationary axisymmetric spacetime
{\boldmath $\eta$} can be chosen as the axial Killing vector with $\omega$ an
arbitrary function of $r$ and $\theta$ in adapted coordinates. For
instance, $\omega$ can be chosen to make the congruence either geodesic
or irrotational. Spatially these will be represent circular orbits. In
cylindrically symmetric spacetimes in addition to the axial Killing vector
we can add on the Killing vector generating $z$-translations with coefficients
as arbitrary functions of $\rho$ in adapted coordinates. Spatially these will
represent helical orbits. In spacetimes admitting other spatial Killing
vectors like De Sitter and G\"{o}del Universes more complicated quasi-Killing
congruences can be generated whose spatial projections would not be
simple curves like circles or helices. Along any particular trajectory
belonging to a quasi-Killing congruence $\omega$ is a constant. With
reference to the congruence in which a trajectory is embedded we may call
such a curve a quasi-Killing trajectory. Of course, if $\omega_{(A)}$ are
constants then {\boldmath $\chi$} defines a Killing trajectory.

\subsection{Frenet-Serret Torsions and Gyroscopic Precession}
The transport law for an observer whose tetrad moves along an arbitrary
world line is written as \cite{mtw*}:
\begin{eqnarray}
\frac{D}{D\tau}\,(e^{a}_{(i)})\,=\,-\Omega^{a}_{\;b}\,e^{b}_{(i)}\label{eq:8}
\end{eqnarray}
where $\Omega$ decomposes into a Fermi-Walker  piece and a spatial rotation
\begin{eqnarray}
\Omega^{ab}\,&=&\,\Omega^{ab}_{\tt (FW)}\,+\,\Omega^{ab}_{\tt (SR)}\nonumber\\
\Omega^{ab}_{\tt (FW)}\,&\equiv&\,a^au^b\,-\,a^bu^a\label{eq:9}\\
\Omega^{ab}_{\tt (SR)}\,&\equiv&\,u_c\omega_d\epsilon^{cdab}\nonumber
\end{eqnarray}
In the above, {\boldmath $ \omega $} is a vector orthogonal to the four
velocity
$u^a$. It is possible to choose the time axis of the tetrad along the
4-velocity of the arbitrary world line consistent with the transport law
equations (16 - 17) and following \cite{mtw*} we restrict to such tetrads.  If
a frame ${\bf f}_{(b)}$ is Fermi-Walker transported along the same world line
the spatial triad of ${\bf e}_{(a)}$ rotates relative to the spatial triad of
${\bf f}_{(a)}$ with
angular velocity $ {\bf \omega}$, {\em i.e.,}
\begin{equation}
\frac{D}{D\tau}\,\left(\bf e_{(\mu)}\,-\,{\bf
f}_{(\mu)}\right)\,=\,\mbox{\boldmath $ \omega $} \times
{\bf e}_{(\mu)}\label{eq:10}
\end{equation}
Comparing the Frenet-Serret equations (\ref{eq:4}) with the transport equations
(\ref{eq:8}-\ref{eq:10})
it is easy to verify that the Frenet-Serret frame rotates with respect to
the Fermi-Walker transported frame by
\begin{equation}
\mbox{\boldmath $ \omega _{\tt (FS)}$}\,=\,\tau_2\,{\bf
e}_{(1)}\,+\,\tau_1\,{\bf e}_{(3)}\label{eq:11}
\end{equation}

The Fermi-Walker frame is physically realized by a system of gyroscopes and
hence the
gyroscopic precession relative to the Frenet-Serret frame - one of the most
natural and intrinsic frames associated with an arbitrary curve - is given
by $-\mbox{\boldmath $\omega_{\tt (FS)}$}$.
\begin{equation}
{\bf \Omega}_{\tt (g)}\,=\,-\mbox{\boldmath $ \omega_{\tt
(FS)}$}\,=\,-(\tau_2\,{\bf
e}_{(1)}\,+\,\tau_1\,{\bf e}_{(3)})\label{eq:12}
\end{equation}
Further, using the Frenet-Serret equation (\ref{eq:4}) one can prove
\begin{equation}
\omega^{a}_{\tt (FS)}\,=\,\tilde{F}^{ab}\,e_{(0)b}\label{eq:13}
\end{equation}
where
\(\tilde{F}^{ab}\,\equiv\,\frac{1}{2\sqrt{-g}}\,\epsilon^{abcd}\,F_{cd}\),
is the dual to $F_{cd}$. We refer to {\boldmath $\omega_{\tt (FS)}$} as
Frenet-Serret rotation. It
should be noted that {\boldmath $\omega_{\tt (FS)}$} is defined along one given
curve. It is not
tied to the existence of a congruence. It gives the rotation of the
Frenet-Serret frame relative to the Fermi-Walker transported frame.

We may mention in passing that from equations (8) and (10) we have
\begin{equation}
\kappa e^{a}_{(1)} = F^{a}_{\;\;b} e^{b}_{(0)}
\end{equation}
which indicates that in analogy with electromagnetism $F^{a}_{\;\;b}
e^{b}_{(0)}$ can be interpreted as the gravielectric field as seen by the
observer with four velocity $e^{a}_{(0)}$. Moreover, the precession
equations (18) and (21) exhibit further suggestive resemblance to the
electromagnetic `spin precession' equations and indicate that $\tilde{F}^{ab}
e_{(0)b}$ is the corresponding gravimagnetic field.

\subsection{Vorticity and Gyroscopic Precession}
Given a trajectory it can be viewed as a member of a suitable chosen congruence
of curves. Associated with a congruence of curves is defined the notion of
vorticity, which geometrically measures the twisting of the congruence.
The gyroscopic precession along a trajectory is related to the vorticity
of the congruence. In this section we shall explore this relation in some
detail.
 It was shown in \cite{hsv74} that  the
Frenet-Serret rotation for a trajectory belonging to the Killing congruence,
is equal to the vorticity of the congruence. Consequently, the gyroscopic
precession for a Killing trajectory is determined by the vorticity of the
Killing congruence.

As we shall show below, in this respect, the quasi-Killing case differs from
the Killing one.
The vorticity of a congruence is defined as
\begin{eqnarray}
\Omega^a\,&\equiv
&\,\frac{1}{2\sqrt{-g}}\,\epsilon^{abcd}\,e_{(0)b}\,e_{(0)c;d}\nonumber\\
&=&\,\frac{1}{2\sqrt{-g}}\,\epsilon^{abcd}\,e_{(0)b}\,\left[
F_{cd}\,+\,e^\psi\,\omega_{(A),d}\,\eta_{(A)c}\right]\label{eq:14}\\
&=&\,\omega^a_{\tt (FS)}\,+\,\tilde{D}^{ab}\,e_{(0)b}\label{eq:15}\\
{\rm where}\;\;
\tilde{D}^{ab}\,&=&\,\frac{1}{2\sqrt{-g}}\,\epsilon^{abcd}\,D_{cd}\nonumber\\
D_{cd}\,&\equiv&\,e^\psi\,\omega_{(A),[d}\,\eta^{(A)}_{c]}
\end{eqnarray}
and antisymmetrisation is defined as
\[
A_{[ab]}\,\equiv \, \frac{1}{2}\,\left( A_{ab}\,-\,A_{ba}\right)
\]

As is well known, physically, vorticity  $\Omega^a$ represents the angular
velocity of the connecting
vector with respect to an orthonormal spatial frame Fermi-Walker
transported along the congruence \cite{jcb92,he}. On the other hand,
Frenet-Serret rotation  $\omega^a_{\tt (FS)}$  represents
precession of the intrinsic Frenet-Serret frame with respect to the
nonrotating Fermi-Walker frame. In general, for example in the quasi-Killing
case, the two are not are the same. Therefore the gyroscopic precession along
a quasi-Killing trajectory differs from the rotation of the connecting vector
of the corresponding quasi-Killing congruence. However,
from eq.(\ref{eq:14}) it follows that if $\omega_{(A)}$
are constants, the congruence $\chi^a$ becomes Killing, $\Omega^a\;=\;
\omega^a_{\tt (FS)}$, and the gyroscopic precession is locked on to the
rotation
of the connecting vector\cite{app}.

The above difference between the two cases, namely  Killing and
quasi-Killing, may also be understood by examining the
Lie derivative of the basis vectors along ${\bf e}_{(0)}$ in the two cases. In
the
Killing case
\begin{equation}
\raisebox{-.25cm}{$\stackrel{\textstyle{\cal L}}{\scriptstyle{{\bf
e}_{(0)}}}$}\,{\bf e}_{(\alpha)}\,=\,
\kappa\,{\bf e}_{(0)}\,\delta^{(1)}_{(\alpha)}
\end{equation}
so that modulo ${\bf e}_{(0)}$ ({\em i.e.,} if one projects normal to
 ${\bf e}_{(0)}$)
the Frenet-Serret frame is Lie dragged along ${\bf e}_{(0)}$. In the quasi
Killing case on the other hand
\begin{equation}
\raisebox{-.25cm}{$\stackrel{\textstyle{\cal L}}
{\scriptstyle{{\bf e}_{(0)}}}$}\,{\bf e}_{(\alpha)}\,=\,
\kappa\,{\bf e}_{(0)}\,\delta^{(1)}_{(\alpha)}\,+\,e^\psi\, \left(
\raisebox{-.25cm}{$\stackrel{\textstyle{\cal
L}}{\scriptstyle{\bf{e_{(\alpha)}}}}$}\,\omega_{(A)}\right )\,\left[ (
\mbox{\boldmath $ \eta_{(A)}$}\,\cdot\,{\bf e}_{(0)})\,{\bf
e}_{(0)}\,-\,\mbox{\boldmath $ \eta_{(A)}$}\right ]
\end{equation}
so that the Frenet-Serret frame is {\em not} Lie dragged along ${\bf e}_{(0)}$.
Recall,
that by definition, the connecting vector is always Lie dragged {\em
i.e., }
\begin{equation}
\raisebox{-.2cm}{$\stackrel{\textstyle{\cal L}}
{\scriptstyle{{\bf e}_{(0)}}}$}\,{\bf c}\,=\,0
\end{equation}

In the following sections, we shall discuss particular examples to
illustrate the application of the above considerations.

\section{Stationary Axially Symmetric Spacetimes}
In this section we specialize to spacetimes which are stationary and
axially symmetric. Such spacetimes have in addition to the timelike
Killing vector {\boldmath $\xi$}, a spacelike Killing vector
{\boldmath $\eta$} with closed orbits. Assuming further
 orthogonal transitivity, in coordinates adapted to the Killing vectors
{\boldmath $\xi$} and {\boldmath $\eta$}, the most general
form of the metric may be written as
\begin{equation}
ds^2\,=\,g_{00}dt^2\,+\,2g_{03}dtd\phi\,+\,g_{33}d\phi^2\,+\,g_{11}dr^2\,+\,
g_{22}d\theta^2\label{eq:16}
\end{equation}
where $g_{ab}$ are functions of $r$ and $\theta$ only.

The contravariant components of the metric may be read off from
\begin{eqnarray}
\left( \frac{\partial}{\partial s}\right
)^2\,&=&\,\frac{g_{33}}{\Delta_3}\,\left( \frac{\partial}{\partial t}\right
)^2\,-2(\frac{g_{03}}{\Delta_3})\,\frac{\partial}{\partial
t}\,\frac{\partial}{\partial\phi}\,+\,\frac{g_{00}}{\Delta_3}\,\left(
\frac{\partial}{\partial\phi} \right)^2 + \nonumber\\
&+&\,\frac{1}{g_{11}}\,\left( \frac{\partial}{\partial r}
\right)^2\,+\,\frac{1}{g_{22}}\,\left(
\frac{\partial}{\partial\theta}\right)^2\label{eq:17}\\
{\rm where
}\,\,\Delta_3\,&\equiv&\,g_{00}\,g_{33}\,-\,g^{2}_{03}\label{eq:18}\\
{\rm and }\,\,\det\,g_{ab}\,&\equiv
&\,g\,=\,\,g_{11}\,g_{22}\Delta_3\label{eq:19}
\end{eqnarray}

After a long but straightforward calculation using
eqs.(\ref{eq:5}-\ref{eq:7}) and eqs.(\ref{eq:16}-\ref{eq:19}) it follows that
along trajectories of the timelike Killing vector {\boldmath $ \xi $} the
Frenet-Serret
invariants are given by
\begin{eqnarray}
\kappa^2\,&=&\,-\frac{1}{4}\,g^{ab}(\ln\,g_{00})_{,a}\,(\ln\,g_{00})_{,b}
\nonumber\label{eq:kitl1}\\
&=&\,-\,\frac{1}{4g_{00}^{2}}\,[g^{11}\,g_{00,1}^{2}\,+\,g^{22}\,g_{00,2}^{2}]
\label{eq:20}\\
\tau^{2}_{1}\,&=&\,\frac{g^{2}_{03}}{4\Delta_3}\,\frac{[g^{ab}\,g_{00,a}\,
(\ln\,\frac{g_{03}}{g_{00}})_{,b}]^2}{[g^{ab}\,g_{00,a}\,g_{00,b}]}
\label{eq:21}\\
\tau^{2}_{2}\,&=&\,\frac{1}{4\Delta_3
g_{11}g_{22}}\,\frac{[g_{00,1}\,g_{03,2}\,-g_{00,2}\,g_{03,1}]^2}
{[g^{ab}\,g_{00,a}\,g_{00,b}]}\label{eq:22}
\end{eqnarray}
In this case the Frenet-Serret basis is given by
\begin{eqnarray}
e^{a}_{(0)}\,&=&\,\frac{1}{\sqrt{g_{00}}}\,(1\,,0\,,0\,,0)\nonumber\\
e^{a}_{(1)}\,&=&\,-\,\frac{1}{2\kappa\,g_{00}}\,(0,\,g^{11}\,g_{00,1},\,g^{22}
\,g_{00,2},\,0)\nonumber\\
e^{a}_{(2)}\,&=&\,\,\frac{1}{\sqrt{g_{00}}\,\sqrt{-\Delta_3}}\,(-g_{03},\,0,\,
0,\,g_{00})\nonumber\\
e^{a}_{(3)}\,&=&\,\,\frac{\sqrt{g^{11}\,g^{22}}}{2\kappa\,g_{00}}\,(0,\,
-g_{00,2},\,g_{00,1},\,0)\label{eq:23}
\end{eqnarray}
Equations (\ref{eq:20}-\ref{eq:23}) completely describe the world line of a
stationary
observer and the precession of a gyroscope carried by him.
$\kappa $ and $\tau_1$ are chosen to be positive and $\tau_2$ taken to
be the positive square root of the right hand side of eq.(\ref{eq:22})
so that ${\bf e}_{(1)},\;
{\bf e}_{(2)},\;{\bf e}_{(3)},\;$ form a right handed triad.
We shall now
apply these formulae to the special case of the Kerr spacetime.\\

\subsection{\bf Kerr spacetime}
The spacetime describing a rotating black hole is the Kerr solution and its
geometry is given by:
\begin{eqnarray}
ds^2\,&=&\,\left(1-\frac{2Mr}{\Sigma}\right)\,dt^2\,-\,\frac{\Sigma}{\Delta}
\,dr^2\,-\,\Sigma
d\theta^2\,+\,\frac{4Mra\,\sin^2\theta}{\Sigma}\,d\phi\,dt\,-\nonumber\\
&-&\,\left(r^2+a^2+\frac{2Mr\,a^2\,\sin^2\theta}{\Sigma}\right)\,\sin^2\theta
d\phi^2 \label{eq:24}\\
{\rm where
}\,\,\Delta\,&\equiv&\,r^2+a^2-2Mr\;\;;\;\;\Sigma\,\equiv\,r^2+a^2\cos^2\theta
\nonumber
\end{eqnarray}

Substituting the above expressions for $g_{ab}$ in
eqs.(\ref{eq:20}-\ref{eq:23})
and simplifying we obtain,
\begin{eqnarray}
\kappa^2\,&=&\,\frac{M^2}{\Sigma^5}\,(\Delta\epsilon^2\,+\,4r^2a^4\,\cos^2
\theta\,\sin^2\theta).\,\,\frac{1}{\left(1-\frac{2Mr}{\Sigma}\right)^2}
\label{eq:ki1}\label{eq:25}\\
\tau^{2}_{1}\,&=&\,\frac{M^2a^2\sin^2\theta}{\Sigma}\,\,\frac{\Delta}
{\left(1-\frac{2Mr}{\Sigma}\right)^2}\,\,\frac{1}{(\Delta\epsilon^2+4r^2a^4
\cos^2\theta\sin^2\theta)}\label{eq:26}\\
\tau^{2}_{2}\,&=&\,\frac{4M^2a^2r^2\cos^2\theta\,\epsilon^2}{\Sigma^3}.\,\,\,
\frac{1}{(\Delta\epsilon^2+4r^2a^4\cos^2\theta\,\sin^2\theta)}\label{eq:27}\\
{\rm where}\,\,\epsilon\,&\equiv&\,r^2-a^2\cos^2\theta\nonumber
\end{eqnarray}

\begin{eqnarray}
e^{a}_{(0)}\,&=&\,\frac{1}{\sqrt{1-\frac{2Mr}{\Sigma}}}\,(1,\,0,\,0,\,0)
\nonumber\\
e^{a}_{(1)}\,&=&\,\frac{1}{\sqrt{\Sigma(\Delta\epsilon^2+4r^2a^4s^2c^2)}}
\,(0,\,\Delta\epsilon,-2ra^2sc,\,0)\nonumber\\
e^{a}_{(2)}\,&=&\,\frac{1}{s\sqrt{\Delta(1-\frac{2Mr}{\Sigma})}}\,
(-\frac{2Mras^2}{\Sigma},\,0,\,0,\,(1-\frac{2Mr}{\Sigma}))\nonumber\\
e^{a}_{(3)}\,&=&\,\frac{1}{\sqrt{\frac{\Sigma}{\Delta}(\Delta\epsilon^2+
4r^2a^4s^2c^2)}}\,(0,\,2ra^2sc,\,\epsilon,\,0)\label{eq:28}
\end{eqnarray}
where $s\,\equiv\,\sin\theta\;\;;\;\;c\,\equiv\,\cos\theta.$

Eqs.(\ref{eq:25}-\ref{eq:28}) show that an observer with fixed spatial
coordinates {\em i.e., } a world line following t-lines, is not only
accelerated
$(\kappa\neq 0)$
but also has an angular velocity relative to the local standards of
nonrotation realized by a set of gyroscopes. This is a manifestation of
the dragging phenomenon in the Kerr spacetime.
For an observer on the equatorial plane $\theta\,=\,\pi/2$
it reduces to
\begin{eqnarray}
\tau_1\,&=&\,\frac{Ma}{r^3}\,\left(1-\frac{2M}{r}\right)^{-1}\label{eq:drag} \\
\tau_2\,&=&\,0
\end{eqnarray}
The bases vectors of the Frenet-Serret frame of the stationary observer(
{\boldmath $\xi $}-lines) are always pointed to the same fixed stars since
they are Lie-dragged along the Killing trajectory. They may be visualized
by a set  of telescopes locked on to the distant stars. They also form the
connecting vectors of the Killing congruence defining stationary
observers. Thus the stationary observers will see the gyroscopes precess
with respect to the distant stars with an angular velocity per unit proper
time given by $-\tau_1$. In sec.4.1.3 we shall discuss the precession of a
gyroscope carried once around a circular orbit as measured by a stationary
observer in his rest frame. To measure the precession {\em relative to a
 gyroscope} carried by the stationary observer the precession due to
 dragging mentioned above needs to be taken into account. Of course,
 for static spacetimes the Frenet-Serret frame of the static observer- locked
on
 to the distant stars- do not precess with respect to the gyroscopes.

It may be worth pointing out that a discussion off the equatorial plane
involves no extra work in this formalism. Thus, we give general expressions
in all cases when one is off the equatorial plane.

\section{Rotating coordinates and Gyroscopic Precession along circular orbits
with constant arbitrary angular speeds}
In section 3, we have obtained $\kappa,\,\tau_1,\,\tau_2$ for an observer
whose world line is along the integral curves of the timelike Killing
vector {\boldmath $\xi$} of a stationary spacetime. Such an observer is at a
fixed
value of $r, \theta$ and $\phi$. In this section we show how
the use of `rotating' coordinates allows one to adapt the expressions of
section 3 to trajectories belonging to a quasi-Killing congruence that
represent observers moving along circular orbits with constant arbitrary
angular speeds. This is in the
spirit of the method used by Rindler and Perlick \cite{rp90}.

Starting from a stationary axially symmetric metric of the form (\ref{eq:16})
adapted to the
Killing vectors {\boldmath $\xi$} and {\boldmath $ \eta $}, we note that
$\mbox{\boldmath $ \xi $}  \,+\,\omega \mbox{\boldmath $ \eta $}$, where
$\omega$ is a constant, is also a Killing vector. A coordinate system
adapted to $\mbox{\boldmath $\xi^\prime $} \,\equiv\,\mbox{\boldmath
$ \xi $} \,+\,\omega \mbox{\boldmath $ \eta $}$ is obtained by a
coordinate transformation
\begin{equation}
\phi\,=\,\phi'\,+\,\omega t'\;\;;\;\;t\,=\,t'
\end{equation}
under which the metric becomes
\begin{eqnarray}
ds^2\,&=&\,g_{0^{\prime}0^{\prime}}\,dt^{\prime
2}\,+\,2g_{0^{\prime}3^{\prime}}\,d\phi^\prime\,dt^\prime\,+\,g_{3^{\prime}
3^{\prime}}\,d\phi^{\prime
2}\,+\,g_{11}\,dr^2\,+\,g_{22}\,d\theta^2\nonumber\\
&&\label{eq:29}\\
{\rm where,}\;\;g_{0^{\prime}0^{\prime}}\,&=&\,g_{00}\,+\,2\omega
g_{03}\,+\,\omega^2g_{33}
\,\equiv\,{\cal A}\label{eq:30}\\
g_{0^{\prime}3^{\prime}}\,&=&\,g_{03}\,+\,\omega g_{33}\,\equiv\,{\cal
B}\label{eq:31}\\
g_{3^{\prime}3^{\prime}}\,&=&\,g_{33}\label{eq:32}
\end{eqnarray}
$\mbox{\boldmath $ \xi^\prime $}\,=\,(1,0,0,0)$ is a Killing vector of this
metric and we can
use eqs.(\ref{eq:20}-\ref{eq:22}) to obtain $\kappa,\,\tau_1$ and $\tau_2$
along this world line.
However, {\boldmath $ \xi^\prime $} corresponds to $\mbox{\boldmath $\xi $}
\,+\,\omega \mbox{\boldmath $\eta $}$ in the unprimed
coordinates so that we can compute $\kappa,\, \tau_1$ and $\tau_2$
along trajectories $\mbox{\boldmath $ \xi $}\,+\,\omega\mbox{\boldmath $\eta$}
$ by replacing $g_{00}\,,g_{03}$ and
$g_{33}$ in eq.(\ref{eq:25}-\ref{eq:27}) by
$g_{0^{\prime}0^{\prime}}\,,g_{0^{\prime}3^{\prime}}$
and $g_{3^{\prime}3^{\prime}}$. More importantly the prescription also
works in cases where $\omega$ is not a constant but only satisfies
$\raisebox{-.25cm}{$\stackrel{\textstyle \cal L}{\scriptstyle
\chi}$}\;\,\omega\,=\,0$. This can be seen by noting that the
expressions for the Frenet-Serret invariants in the quasi-Killing case do
not involve derivatives of $\omega$. One can also check explicitly that
the same expressions for $\kappa\,, \tau_1, \tau_2$ obtains whether one starts
from $\mbox{\boldmath $\xi$}\,+\,\omega\mbox{\boldmath $\eta$}$ or if one uses
 the expressions for {\boldmath $\xi$} and
replaces $g_{ab}$ by $g_{a^{\prime}b^{\prime}}$ treating $\omega$ as a
constant.
Thus along trajectories of $\mbox{\boldmath $\xi$}\,+\,\omega\mbox{\boldmath
 $ \eta $}$ we obtain
\begin{eqnarray}
\kappa^2\,&=&\,-\,\frac{1}{4}\,\frac{(g^{11}{\cal A}_{(1)}^2+
g^{22}{\cal A}_{(2)}^2)}{{\cal A}^2}\label{eq:33}\\
\tau^{2}_{1}\,&=&\,\left(\frac{{\cal B}^2}{4\Delta_3(g^{11}{\cal
A}_{(1)}^2+g^{22}
{\cal A}_{(2)}^2)}\right).\nonumber\\
&.&\left(\frac{g^{11}{\cal A}_{(1)}{\cal B}_{(1)}+g^{22}{\cal
A}_{(2)}{\cal B}_{(2)}}{{\cal B}} -\frac{g^{11}{\cal A}_{(1)}^2+g^{22}{\cal
A}_{(2)}^2}{{\cal A}}\right)^2\label{eq:34}\\
\tau^{2}_{2}\,&=&\,\frac{g^{11}\,g^{22}({\cal A}_{(1)}{\cal B}_{(2)}-
{\cal A}_{(2)}{\cal B}_{(1)})^2}{4\Delta_3\,
(g^{11}{\cal A}_{(1)}^2+g^{22}{\cal A}_{(2)}^2)}\label{eq:35}\\
{\rm where}\;\;{\cal A}_{(a)}\,&\equiv &\,g_{00,a}+2\omega
g_{03,a}+\omega^2g_{33,a}\;;\;a=1,2.\label{eq:36}\\
{\cal B}_{(b)}\,&\equiv &\,g_{03,b}+\omega g_{33,b}\;;\;b=1,2.\label{eq:37}
\end{eqnarray}

To obtain the Frenet-Serret tetrad associated with {\boldmath $\chi$} we recall
that
with respect to the primed coordinates {\boldmath $\chi $} is like
{\boldmath $\xi $}. Thus the Frenet-Serret
tetrad in the primed coordinates are obtained by replacing $g_{ab}\;$ by
$g_{a'b'}\;$ in eq.(\ref{eq:23}).  The components relative to the original
unprimed coordinates  are obtained
via a vector transformation and finally we find:
\begin{eqnarray}
e^{a}_{(0)}\,&=&\,\frac{1}{\sqrt{\cal A}}\,(1\,,0\,,0\,,\omega)\nonumber\\
e^{a}_{(1)}\,&=&\,-\,\frac{1}{2\kappa\,{\cal A}}(0,\,g^{11}\,{\cal A}_{(1)},
\,g^{22}\,{\cal A}_{(2)},\,0)\nonumber\\
e^{a}_{(2)}\,&=&\,\,\frac{1}{\sqrt{{\cal A}}\,\sqrt{-\Delta_3}}\,
({\cal B},\,0,\,0,\,-{\cal C})\nonumber\\
e^{a}_{(3)}\,&=&\,\,\frac{\sqrt{g^{11}\,g^{22}}}{2\kappa\,{\cal A}}
(0,\,-{\cal A}_{(2)},\,{\cal A}_{(1)},\,0)\label{eq:38}\\
{\rm where}\,\,{\cal C}\,&\equiv&\,g_{00}+\omega g_{03}\nonumber
\end{eqnarray}

For later comparison we write down the dual bases below. They are given by,
\begin{eqnarray}
{\bf\omega}^{(0)}\,&=&\,\frac{1}{\sqrt{{\cal A}}}[{\cal C}{\bf dt}\,-
\,{\cal B}{\bf d\phi}]\nonumber\\
{\bf\omega}^{(1)}\,&=&\,\frac{1}{2\kappa{\cal A}}[{\cal A}_{(1)}{\bf dr}\,+\,
{\cal A}_{(2)}{\bf d\theta}]\nonumber\\
{\bf\omega}^{(2)}\,&=&\,\sqrt{\frac{-\Delta_3}{{\cal A}}}[\omega {\bf dt}\,-\,
{\bf d\phi}]\nonumber\\
{\bf\omega}^{(3)}\,&=&\,\frac{1}{2\kappa{\cal A}}[\sqrt{\frac{g_{22}}{g_{11}}}
{\cal A}_{(2)}{\bf dr}\,-\,\sqrt{\frac{g_{11}}{g_{22}}} {\cal A}_{(1)} {\bf
d\theta}]\label{eq:39}
\end{eqnarray}

We next apply the above formulas to various special cases and
retrieve well known gyroscopic precessions. Many of our formulas are more
general in that they are not confined to the equatorial plane but valid
off it. The formulas are moreover demonstrated in a unified framework.

It should be recalled that in the above $\omega$ can be any arbitrary
function of $r$ and $\theta$. This allows one to discuss precession along a
 geodesic (where $\omega$ is not a constant) by chosing $\omega$ such
that $\kappa\,=\,0$. Although the Frenet-Serret equations are reduced to only
one,
$\tau_1$ and $\tau_2$ are defined through $\omega$, ${\bf
e}_{(2)},\,$ and ${\bf e}_{(3)}$ by  considering the geodesic as a
particular limit of the congruence obtained by keeping
$\omega$ constant
corresponding to the geodetic value. Schwarzschild metric corresponds to
$a\,=\,0$ while flat spacetime to $M\,=\,a\,=\,0$. In addition if
trajectories are  confined to the equatorial plane then
$\theta\,=\,\pi/2$.
\subsection{Kerr Black Hole}
\subsubsection{The general case}
The general procedure outlined above may be applied to obtain the
acceleration and gyroscopic precession in the case of an observer
following a quasi-Killing trajectory in the Kerr spacetime. A straightforward
computation yields:
\begin{eqnarray}
\kappa^2\,&=&\,\frac{{\cal K}_1}{\Sigma{\cal K}_2 }\label{eq:40}\\
\tau^{2}_{1}\,&=&\,\frac{\Delta}{\Sigma}\,\frac{{\cal K}_3 }{{\cal K}_1{\cal
K}_2 }\,s^2\label{eq:41}\\
\tau^{2}_{2}\,&=&\,\frac{M^2 {\cal K}_4 }{\Sigma^5 {\cal
K}_1}\,c^2\label{eq:42}\\
{\rm where}\;\;
{\cal K}_1&=&\Delta\left[\frac{M\epsilon}{\Sigma^2}(1-a\omega
s^2)^2-rs^2\omega^2\right]^2+\nonumber\\
        &+&c^2s^2\left[\frac{2Mr}{\Sigma^2}\{(r^2+a^2)\omega
-a\}^2+\Delta\omega^2\right]^2\label{eq:43}\\
{\cal K}_2&=&\left[1-(r^2+a^2)s^2\omega^2-\frac{2Mr}{\Sigma}(1-a\omega
s^2)^2\right]^2\label{eq:44}\\
{\cal K}_3&=&\left[\{\frac{M\epsilon}{\Sigma^2}(1-a\omega
s^2)-rs^2\omega^2\}\cdot\right.\nonumber\\
          &.&\left.[r\omega-\frac{2Mr^2}{\Sigma}(1-a\omega
s^2)\omega-\frac{M\epsilon}{\Sigma^2}(1-a\omega s^2)\{(r^2+a^2)\omega
-a\}]+\right.\nonumber\\
          &+&\left.c^2\{\frac{2Mra}{\Sigma^2}(1-a\omega
s^2)^2-\omega\}\{\frac{2Mr}{\Sigma^2}((r^2+a^2)\omega
-a)^2+\Delta\omega^2\}\right]\nonumber\\
& &\\
{\cal K}_4&=&\left[\frac{2Mr\epsilon a}{\Sigma}(1-a\omega s^2)^2-\epsilon\omega
(r^2+a^2) (1-a\omega s^2)+\right.\nonumber\\
&+&\left.2as^2\omega r^2\{(r^2+a^2)\omega -a\}\right]^2\label{eq:45}
\end{eqnarray}

The bases are given by eq.(\ref{eq:38}) with
${\cal A},\,\,{\cal A}_{(1)},\,\,{\cal A}_{(2)},\,\,{\cal B}$ and ${\cal C}$
given by:
\begin{eqnarray}
{\cal A}&=&1-\omega^2s^2(r^2+a^2)-\frac{2Mr}{\Sigma}(1-a\omega
s^2)^2\nonumber\\
{\cal A}_{(1)}&=&\frac{2M\epsilon}{\Sigma^2}(1-a\omega
s^2)^2-2r\omega^2s^2\nonumber\\
{\cal A}_{(2)}&=&-2cs[\Delta\omega^2+\frac{2Mr}{\Sigma^2}\{(r^2+a^2)\omega
-a\}^2]\nonumber\\
{\cal B}&=&\frac{2Mras^2}{\Sigma}(1-a\omega s^2)-(r^2+a^2)\omega s^2\nonumber\\
{\cal C}&=&1-\frac{2Mr}{\Sigma}(1-a\omega s^2)\label{eq:46}
\end{eqnarray}
As explained earlier for $\omega\,=\,0$ eqs.(\ref{eq:40}-\ref{eq:45}) reduce to
eqs.(\ref{eq:25}-\ref{eq:28}) for motion along {\boldmath $\xi$}, the global
timelike Killing vector defining a stationary observer.
\subsubsection{The equatorial plane $\theta\,=\,\pi/2$.}
On the equatorial plane the above expressions reduce to:
\begin{eqnarray}
\kappa^2\,&=&\,\frac{\Delta M^2}{r^6}\,\frac{\left[ (a\omega\,-\,1)^2\,-\,
\frac{r^3\omega^2}{M}\,\right]^2}{\left[1\,-\,(r^2\,+\,a^2)\omega^2\,-\,
\frac{2M\,(a\omega-1)^2}{r}\,\right]^2}\\
\tau^{2}_{1}\,&=&\,\frac{1}{r^2}\,\frac{\left[ \frac{Ma}{r^2}\,-\,
\left(\frac{(r^2\,+\,2a^2)\,M}{r^2}\,-\,r(1\,-\,\frac{2M}{r})\right)\omega\,+
\,\frac{Ma(3r^2\,+\,a^2)\omega^2}{r^2}\,\right]^2}{\left[1-\,(r^2\,+\,a^2)
\omega^2\,-\,\frac{2M(a\omega-1)^2}{r}\right]^2}\\
\tau^{2}_{2}\,&=&\,0.
\end{eqnarray}
The bases are given by eqs.(\ref{eq:38}-\ref{eq:46}) with $s=1$, $c=0$.

We note that the gyroscopic precession is about ${\bf e}_{(3)}$ which is normal
to the orbital plane and the precession frequency is given by $\tau_1$
as above.
\subsubsection{Geodesic motion and Schiff precession}
Along a geodesic, $\kappa\,=\,0$ whence
\begin{equation}
\omega^{-1}\,=\,a\,\pm\,\sqrt{\frac{r^3}{M}}\label{eq:sch}
\end{equation}
  This yields the Keplerian frequency in the Kerr case and the Frenet-Serret
invariant for motion along this geodesic is
\begin{equation}
\tau^{2}_{1}\,=\,\frac{M}{r^3}
\end{equation}
In eq.(\ref{eq:sch}) the $+(-)$ signs correspond to
co-rotating(counter-rotating)
 orbits. The range of values of $ r$ for which these orbits are timelike have
 been discussed in sec.2.2 of \cite{rp90}. Their analysis shows that the range
  of  $ r$ for which counter-rotating orbits are timelike requires that the
absolute value
 of $\,a\,$ be less than the modulus of $\sqrt{\frac{r^3}{M}}$.
It should be noted that as $\omega$ approaches the Keplerian value the
combination ${\cal A}_{(1)}/\kappa{\cal A}$ is still well-defined leading
to ${\bf e}_{(1)}$ and ${\bf e}_{(3)}$ independent of $\omega$. This allows us
to
extract the geodetic case as a special instance of our more general
motion.

The gyroscopic precession frequency, $\,\mp |\tau_1 |,\,$ is thus
$\,\mp\sqrt{\frac{M}{r^3}}\,.$  This precession is about $\,{\bf e}_{(3)}\,$
which coincides with the {\em z}-direction. The orbiting (corotating)
observer measures
precession relative to  $\,{\bf e}_{(1)}\,$ which coincides with her radius
vector which rotates with angular velocity  $\,\omega\,$
 given by eq.(\ref{eq:sch}).  The precession angle per unit proper time as
 computed in the rotating coordinates is therefore
\begin{equation}
\Delta\phi'\,=\,\mp\frac{M}{r^3}\,\sqrt{g_{0'0'}}\,\frac{2\pi}{|\omega|}
\end{equation}
where $\omega$ is the angular frequency of rotation per unit coordinate
time. This agrees with the results of Rindler and Perlick\cite{rp90}. The
baseline with respect to which the precession is calculated in the rotating
coordinates by them coincides with the Frenet-Serret vector
$\,{\bf e}_{(1)}\,$ on the equatorial plane. Consequently, it leads to the
same precession angle $\,\Delta\phi'\,$. In order to compute the
precession relative to a stationary geometry( `frame of fixed stars') we
need to subtract from the precession at the end of one revolution the amount
through which $\,{\bf e}_{(1)}\,$ has rotated with respect to the stationary
observer, namely  $\,2\pi\,$ radians. Following this procedure we arrive
at the gyroscopic precession in the Kerr spacetime:
\begin{equation}
\Delta\phi\,=\,\mp 2\pi\left[\left(1-\frac{3M}{r}\pm
2a\sqrt{\frac{M}{r^3}}\right)^{\frac{1}{2}}-1\right].
\end{equation}
In the linear approximation this reduces to the Schiff precession.
This agrees with the standard results quoted in literature
including ref.\cite{rp90}.

As discussed earlier in sec. 3.1 one may want to compute the precession of
the orbiting gyroscope with respect to the fiducial gyroscope of the stationary
observer. In one revolution of the orbiting gyroscope the latter precesses
due to dragging by an amount
\begin{equation}
\Delta\phi_{\tt (drag)}\,=\,(-\tau_1)\,\sqrt{g_{00}}\,\frac{2\pi}{|\omega |}
\end{equation}
where $\tau_1$ is given by eq.(\ref{eq:drag}). This leads to\cite{err}
\begin{equation}
\Delta\phi_{\tt (drag)}\,=\,- \frac{2\pi Ma}{r^3}
\left(\sqrt{\frac{r^3}{M}}\,\,\pm\,\,a\right)
\left(1-\frac{2M}{r}\right)^{-\frac{1}{2}}\label{eq:drpr}
\end{equation}
\subsection {Schwarzschild black hole}
The Schwarzschild metric may be obtained from the Kerr metric by setting
$a=0$. Correspondingly the most general case of gyroscopic precession
follows from the Kerr expression for $a=0$.
\subsubsection{General Schwarzschild case}
The  $a\,=\,0$ limit of eqs.(\ref{eq:40}-\ref{eq:46}) yields:
\begin{eqnarray}
\kappa^2\,&=&\,r^2\,\frac{[(1-\frac{2M}{r})(\frac{M}{r^3}\,-\,\omega^2
s^2)^2\,+\,\omega^4s^2c^2]}{(1-\frac{2M}{r}\,-\,r^2\omega^2s^2)^2}\label{eq:47}
\\
\tau^{2}_{1}\,&=&\,\omega^2 s^2\,\frac{(1-\frac{2M}{r})
[(\frac{M}{r^3}-\omega^2s^2)(1-\frac{3M}{r})-\omega^2c^2]^2}
{(1-\frac{2M}{r}-r^2\omega^2s^2)^2[(1-\frac{2M}{r})(\frac{M}{r^3}-\omega^2s^2
)^2+\omega^4s^2c^2]} \label{eq:48}\\
\tau^{2}_{2}\,&=&\,\frac{\omega^2 M^2c^2}{r^6[(1-\frac{2M}{r})(\frac{M}{r^3}-
\omega^2s^2)^2+\omega^4s^2c^2]}\label{eq:49}
\end{eqnarray}
The Frenet-Serret frame is given by:
\begin{eqnarray}
e^{a}_{(0)}&=&\frac{1}{\sqrt{1-\frac{2M}{r}-r^2\omega^2s^2}}\,\,(1, 0, 0,
\omega)\nonumber\\
e^{a}_{(1)}&=&\frac{1}{[(1-\frac{2M}{r})(\frac{M}{r^3}-\omega^2s^2)^2+
\omega^4s^2c^2]^{1/2}}\,(0,(1-\frac{2M}{r})(\frac{M}{r^3}-\omega^2s^2),\,
\frac{\omega^2cs}{r},\,0)\nonumber\\
e^{a}_{(2)}&=&\frac{1}{rs\sqrt{(1-\frac{2M}{r})(1-\frac{2M}{r}-
\omega^2r^2s^2)}}\,(\omega
r^2s^2, 0, 0, -(1-\frac{2M}{r}))\nonumber\\
e^{a}_{(3)}&=&\frac{\sqrt{1-\frac{2M}{r}}}{r[(1-\frac{2m}{r})(\frac{M}{r^3}-
\omega^2s^2)^2+\omega^4c^2s^2]^{1/2}}\,\,(0,\,csr\omega^2,\,
\frac{M}{r^3}-\omega^2s^2,\,0) \label{eq:50}
\end{eqnarray}
\subsubsection{The equatorial plane}
Most commonly, the precession is computed for orbits in the equatorial
plane for which $\theta =\pi/2$.
Eqs.(\ref{eq:47}-\ref{eq:49}) reduce when $\theta=\frac{\pi}{2}$ to
\begin{eqnarray}
\kappa^2\,&=&\,\frac{r^2(1-\frac{2M}{r})(\frac{M}{r^3}-\omega^2)^2}
{(1-\frac{2M}{r}-r^2\omega^2)^2}\\
\tau^{2}_{1}\,&=&\,\omega^2\,\frac{(1-\frac{3M}{r})^2}{(1-\frac{2M}{r}
-r^2\omega^2)^2}\\
\tau^{2}_{2}\,&=&\,0
\end{eqnarray}
The bases vectors of the Frenet-Serret frame obtain by inserting $s=1,
c=0$ in eq.(\ref{eq:50}). The $\omega$-independence of ${\bf e}_{(1)}$ and
${\bf e}_{(3)}$ mentioned
earlier may be noted more transparently in this instance.

The gyroscopic precession in this case is
\begin{equation}
\tau_1\,=\,\omega(1-\frac{3M}{r})(1-\frac{2M}{r}-r^2\omega^2)^{-1}
\end{equation}
and
\begin{equation}
\Delta\phi\,=\,-2\pi\,[(1-\frac{3M}{r})(1-\frac{2M}{r}-r^2\omega^2)^{-1/2}\,-1]
\end{equation}
\subsubsection{Fokker-De Sitter precession}
Along a geodesic $\kappa\,=\,0$ so that we recover the Keplerian frequency
\[\omega^2\,=\,\frac{M}{r^3}.\]
In this case,
\[\tau^{2}_{1}\,=\,\omega^2\]
so that the orbital gyroscopic
precession frequency is $\omega$, the same as the angular speed $\omega$.
In one orbital revolution, the
gyroscope rotates by
\begin{equation}
\Delta\phi\,=\,-2\pi\,[(1-\frac{3M}{r})^{1/2}\,-1]
\end{equation}
\subsection{Minkowski spacetime}
\subsubsection{The general case}
This corresponds to $M=0$ in eqs.(\ref{eq:47}-\ref{eq:49}) whence
\begin{eqnarray}
\kappa^2\,&=&\,\frac{r^2\omega^4s^2}{(1-r^2\omega^2 s^2)^2}\\
\tau^{2}_{1}\,&=&\,\frac{\omega^2}{(1-r^2\omega^2s^2)^2}\\
\tau^{2}_{2}\,&=&\,0
\end{eqnarray}
while eqs.(\ref{eq:50}) reduce to
\begin{eqnarray}
e^{a}_{(0)}&=&\frac{1}{\sqrt{1-r^2s^2\omega^2}}(1, 0, 0, \omega)\nonumber \\
e^{a}_{(1)}&=&(0, -s, \frac{c}{r}, 0)\nonumber \\
e^{a}_{(2)}&=&\frac{1}{rs\sqrt{1-\omega^2r^2s^2}}(\omega r^2s^2, 0, 0,
-1)\nonumber \\
e^{a}_{(3)}&=&(0, c, -\frac{s}{r}, 0)
\end{eqnarray}
Note that $\tau_2$ vanishes identically. Therefore, the precession is
about the normal to the orbital plane as should be expected from the symmetry
of the situation.
\subsubsection{Thomas precession}
The above expressions reduce on the $\theta\,=\,\pi/2$ plane to:
\begin{eqnarray}
\kappa^2\,&=&\,\frac{r^2\omega^4}{(1-r^2\omega^2)^2}\\
\tau^{2}_{1}\,&=&\,\frac{\omega^2}{(1-r^2\omega^2)^2}\\
\tau^{2}_{2}\,&=&\,0
\end{eqnarray}
leading to the familiar expression for Thomas precession :
\begin{eqnarray}
\Delta\phi\,=\,-2\pi\,[(1-r^2\omega^2)^{-1/2}\,-\,1]
\end{eqnarray}
As expected the `Keplerian' analog is $\omega\,=\,0$ in which case there
is no precession at all!

In the above sections we have shown how the general Frenet-Serret
formalism can be adapted to retrieve the results discussed in {\em e.g.}
ref. \cite{rp90}.
Motions more general than geodetic or confined to the equatorial plane are easy
to
include and formulas corresponding to these cases have also been exhibited.
\subsection{Globally Hypersurface Orthogonal Stationary Trajectories (GHOSTs)}
The Kerr spacetime admits an important congruence which conforms to our
definition of quasi-Killing vector fields. Observers adapted to this
congruence have been called Locally Non-Rotating Observers (LNROs) or Zero
Angular Momentum Observers (ZAMOs). Considerable insight into the physical
significance of phenomena occurring in the Kerr spacetime is gained by
studying them with reference to the above observers \cite{mtw}.
In the broader context of orthogonal
transitivity, it was shown \cite{gsv75} that
this congruence consists of what we may term as Globally Hypersurface
Orthogonal Stationary Trajectories or GHOSTs, with t = constant being the
hypersurfaces to which they are orthogonal. Therefore, the vorticity of
the congruence identically vanishes, so that the connecting vector between
two adjacent trajectories does not precess with respect to the
Fermi-Walker transported gyroscopes.

The quasi-Killing vector corresponding to the LNRO/ZAMO/GHOST is defined by
\begin{eqnarray}
{\bf \chi}\,&=&\,{\bf \xi}\,+\,\omega{\bf\eta},\,\,\\
\omega\,&=&\,-\,\frac{\xi\cdot\eta}{\eta\cdot\eta}\,=\,-\frac{g_{03}}{g_{33}}
\end{eqnarray}
We note that {\boldmath $\chi$} is timelike down to the event horizon on which
it
becomes null.

As mentioned earlier, the vorticity of this congruence is zero, so that
\begin{eqnarray}
 \Omega^{a}_{\sc\tt
GHOST}\,=\,0\,&=&\,\frac{1}{2\sqrt{-g}}\,\epsilon^{abcd}\,u_bu_{c;d}\nonumber\\
&=&\,\frac{1}{2\sqrt{-g}}\,\epsilon^{abcd}\,u_b\,F_{cd}\,+\,\frac{e^{2\psi}}
{2\sqrt{-g}}\,\epsilon^{abcd}\,\xi_b\eta_c\omega_{,d}\\
                      &=&\,\omega^{a}_{\tt (FS)}\,+\,\Omega^{a}_{\tt (prec)}\,.
\end{eqnarray}
Consequently, the connecting vector between two neighbouring trajectories
belonging to this congruence does not precess relative to the Fermi-Walker
transported gyroscopes. Further,{\boldmath $\omega_{\tt (FS)}$} is the
precession of the Frenet-Serret
frame with respect to the gyroscopes.  Therefore, precession of the
gyroscopes relative to the Frenet-Serret frame is given by $-\mbox{\boldmath
$\omega $}_{\tt (FS)}$ which is
equal to ${\bf\Omega}_{\tt (prec)}$ {\em only for the irrotational congruence}.
The
expression for {\boldmath $\omega_{\tt (FS)}$} is the same as one would obtain
if the
particular trajectory is treated as a member of the {\it Killing
congruence} obtained by taking $\omega$ of the trajectory as a constant
for the entire congruence. This also means that the Frenet-Serret frame is
rigidly
attached to the connecting vector associated with this {\it Killing
congruence}. The expression $\Omega_{\tt (prec)}$ is exactly the same as given
in \cite{mtw} which is the precession of the gyroscope with respect to the
locally
orthogonal triad of \cite{mtw} adapted to the irrotational congruence. To sum
up,
because of the vanishing vorticity, the connecting vector between adjacent
observers following the irrotational congruence is locked on to the
inertial system of gyroscopes and does not precess with respect to the latter.
However, gyroscopes do precess relative to the Frenet-Serret frame and the
latter are not
inertial. This precession frequency in this case is given by two equivalent
expressions
one of them involving derivatives of $\omega$. As we shall show later, the
Frenet-Serret triad coincides with the triad defined in [1] on the equatorial
plane
but differs by a constant spatial rotation for $\theta \neq
\frac{\pi}{2}$.

We may note in passing that in the case of a LNRO Thorne and MacDonald
\cite{tm82}
write down the Fermi-Walker time derivative of any vector orthogonal to
$u^a$ as
\begin{eqnarray}
D_\tau\,{\bf M}\,&=&\,\alpha^{-1}[{\raisebox{-.25cm}
{$\stackrel{\textstyle{\cal L}}{\scriptstyle{\bf t}}$}\;}{\bf M}\,+\,\omega
{\raisebox{-.25cm}{$\stackrel{\textstyle{\cal L}}{\scriptstyle{\bf m}}$}\;}{\bf
M}\,
+\,\frac{1}{2}\,({\bf m}\,\times\,\nabla\omega)\,\times\,{\bf M}]\\
{\rm where}\,\,{\bf m}\,&\equiv &\,\mbox{\boldmath $\eta $} \;\;{\rm and}
\;\;\alpha\,=\,\sqrt{\mbox{\boldmath $\chi $}\cdot\mbox{\boldmath $ \chi$}}
\end{eqnarray}
If ${\bf M}$ is taken as the Frenet-Serret spatial triad then
$\raisebox{-.25cm}{$\stackrel{\textstyle{\cal L}}{\scriptstyle {\bf t}}$}\;{\bf
M}\,=\,
\raisebox{-.25cm}{$\stackrel{\textstyle{\cal L}}{\scriptstyle {\bf m}}$}\;{\bf
M}\,=\,0 $
and the formula reduces to the precession of the triad relative to the
gyroscopes
with frequency $\frac{1}{2}\,({\bf m}\,\times\,\nabla\omega)$ which is an
equivalent
form of our expression for the precession frequency.

We now consider the congruence within the framework of our formalism. It
should be clear from our discussion of the quasi-Killing congruence and
the section on the use of rotating coordinates that all our formulas are
applicable when referring to {\em a particular fixed curve} of any congruence -
in
particular the GHOST. Consequently, to calculate the precession of
gyroscopes relative to a GHOST we proceed exactly as in the earlier
cases and use for $\omega$ the expression appropriate to a GHOST {\it
i.e.,} $\omega\,=\,-g_{03}/g_{33}$.
Since none of our formulas involve differentiation of $\omega$ the
same expressions eqs.(\ref{eq:33}-\ref{eq:38}) give the formula for precession
of a gyroscope relative to
the Frenet-Serret frame of the GHOST. Thus we obtain
\begin{eqnarray}
\kappa^2\,&=&\,-\,\frac{1}{4}\,\frac{g^{ab}(\frac{\Delta_3}{g_{33}})_{,a}
(\frac{\Delta_3}{g_{33}})_{,b}}{(\frac{\Delta_3}{g_{33}})^2} \\
\tau^{2}_{1}\,&=&\,\frac{g^{2}_{33}}{4\Delta_3}\,\frac{[g^{ab}\,
(\frac{\Delta_3}{g_{33}})_{,a}\,
(\frac{g_{03}}{g_{33}})_{,b}]^2}{g^{ab}\,(\frac{\Delta_3}{g_{33}})_{,a}\,
(\frac{\Delta_3}{g_{33}})_{,b}}\\
\tau^{2}_{2}\,&=&\,\frac{g^{2}_{33}}{4\Delta_3g_{11}g_{22}}\,\frac
{[(\frac{\Delta_3}{g_{33}})_{,1}(\frac{g_{03}}{g_{33}})_{,2}\,-\,(
\frac{\Delta_3}{g_{33}})_{,2}(\frac{g_{03}}{g_{33}})_{,1}]^2}{g^{ab}\,
(\frac{\Delta_3}
{g_{33}})_{,a}(\frac{\Delta_3}{g_{33}})_{,b}}
\end{eqnarray}
\begin{eqnarray}
e^{a}_{(0)}&=&(\frac{g_{33}}{\Delta_3})^{1/2}(1, 0, 0,
-\frac{g_{03}}{g_{33}})\nonumber\\
e^{a}_{(1)}&=&-\frac{1}{\sqrt{g^{11}(\frac{\Delta_3}{g_{33}})^{2}_{,1}+g^{22}
(\frac{\Delta_3}{g_{33}})^{2}_{,2}}}
(0,
g^{11}(\frac{\Delta_3}{g_{33}})_{,1},\,\,g^{22}(\frac{\Delta_3}{g_{33}})_{,2}
,0)\nonumber\\
e^{a}_{(2)}&=&-(0, 0, 0, \frac{1}{\sqrt{-g_{33}}})\nonumber\\
e^{a}_{(3)}&=&\frac{\sqrt{g^{11}g^{22}}}{\sqrt{g^{11}(\frac{\Delta_3}{g_{33}})
^{2}_{,1}+g^{22}(\frac{\Delta_3}{g_{33}})^{2}_{,2}}}
(0, - (\frac{\Delta_3}{g_{33}})_{,2}, (\frac{\Delta_3}{g_{33}})_{,1},0)
\end{eqnarray}
These observers accelerate $(\kappa\ne 0)$ and their Frenet-Serret frames
precess
with respect to the gyroscopes $(\tau_1,\tau_2 \ne \,0)$.

The above expression can be calculated explicitly for the Kerr solution.
This gives,
\begin{eqnarray}
\kappa^{2}&=&\frac{M^2[{\cal L}^{2}_{1}+\Delta{\cal
L}^{2}_{2}s^2c^2]}{\Sigma^3\Delta{\cal L}^{2}_{3}}\\
\tau^{2}_{1}&=&\frac{M^2a^2s^2[\Sigma{\cal L}_1{\cal L}_4+2ra^2{\cal L}_2\Delta
s^2c^2]^2}{\Sigma^5{\cal L}^{2}_{3}[{\cal L}^{2}_{1}+{\cal L}^{2}_{2}\Delta
s^2c^2]}\\
\tau^{2}_{2}&=&\frac{4M^2r^2a^6s^4c^2\Delta
[{\cal L}_1+(r^2+a^2){\cal L}_4]^2}{\Sigma^5{\cal L}^{2}_{3}[{\cal
L}^{2}_{1}+{\cal L}^{2}_{2}\Delta s^2c^2]}\\
e^{a}_{(0)}&=&\sqrt{\frac{{\cal L}_3}{\Delta}}(1, 0, 0,
\frac{2Mra}{\Sigma{\cal L}_3})\nonumber\\
e^{a}_{(1)}&=&\frac{M}{\kappa\Sigma^2{\cal L}_3}(0, {\cal L}_1, {\cal L}_2sc,
0)\nonumber\\
e^{a}_{(2)}&=&-(0, 0, 0, \sqrt{{\cal L}_3}s)\nonumber\\
e^{a}_{(3)}&=&\frac{M}{\kappa\Sigma^2{\cal L}_3}(0,
-\sqrt{\Delta}{\cal L}_2sc, \frac{{\cal L}_1}{\sqrt{\Delta}}, 0)\\
{\rm where}\,\,\,{\cal L}_1&=&r^4-a^4+\frac{2a^2s^2r^2\Delta}{\Sigma}\\
{\cal L}_2&=&\frac{2ra^2(r^2+a^2)}{\Sigma}\\
{\cal L}_3&=&r^2+a^2+\frac{2Mra^2s^2}{\Sigma}\\
{\cal L}_4&=&2r^2+(r^2+a^2)\frac{\epsilon}{\Sigma}
\end{eqnarray}
Specializing eqs.(\ref{eq:39}) to the GHOST it is easy to see after a
little computation that the Frenet-Serret
frame coincides with the LNRO frame in \cite{mtw} if $\theta\,=\,\pi/2$. The
Frenet-Serret frame is in general  oriented so that ${\bf e}_{(1)}$ is
along the direction of the acceleration which  is
not along the r-direction, if the orbit is not confined to the equatorial
plane.
\subsection{ De Sitter Universe}
We next apply the formulas to the case of the De Sitter
universe whose metric we take in the form
\begin{eqnarray}
ds^2\,=\,(1-\frac{r^2}{\alpha^2})\,dt^2\,-\,(1-\frac{r^2}{\alpha^2})^{-1}dr^2
\,-\,r^2d\theta^2\,-\,r^2\sin^2\theta\,d\phi^2
\end{eqnarray}
Along trajectories of $\mbox{\boldmath $\xi $}\,+\,\omega\mbox{\boldmath $
\eta $}$ in this case we obtain:
\begin{eqnarray}
\kappa^2\,&=&\,r^2\,\frac{[(1-\frac{r^2}{\alpha^2})(\frac{1}{\alpha^2}\,+\,
\omega^2s^2)^2\,+\,\omega^4s^2c^2]}{(1-\frac{r^2}{\alpha^2}\,-\,r^2\omega^2s^2)
^2}\\
\tau^{2}_{1}\,&=&\,\frac{\omega^2s^2(1-\frac{r^2}{\alpha^2})(\frac{1}{\alpha^2}
+\omega^2)^2}{(1-\frac{r^2}{\alpha^2}-r^2\omega^2s^2)^2
[(1-\frac{r^2}{\alpha^2})(\frac{1}{\alpha^2}+\omega^2s^2)^2+\omega^4s^2c^2]}\\
\tau^{2}_{2}\,&=&\,\frac{\omega^2c^2}{\alpha^4[(1-\frac{r^2}{\alpha^2})
(\frac{1}{\alpha^2}+\omega^2s^2)^2+\omega^4s^2c^2]}
\end{eqnarray}
\begin{eqnarray}
e^{a}_{(0)}&=&\frac{1}{\sqrt{{\cal S}_1}}(1,\; 0,\; 0,\; \omega)\nonumber\\
e^{a}_{(1)}&=&\frac{1}{\sqrt{{\cal S}_2}}(0,\;(1-\frac{r^2}{\alpha^2})
(\frac{1}{\alpha^2}+\omega^2s^2),\;
\frac{-\omega^2sc}{r},\; 0)\nonumber\\
e^{a}_{(2)}&=&\frac{1}{\sqrt{{\cal S}_1}}\left(\frac{\omega rs}{\sqrt{1-
\frac{r^2}{\alpha^2}}}\,,\; 0,\; 0,\;
-\frac{\sqrt{1-\frac{r^2}{\alpha^2}}}{rs}\right)\nonumber\\
e^{a}_{(3)}&=&\frac{\sqrt{1-\frac{r^2}{\alpha^2}}}{\sqrt{{\cal S}_2}}(0,\;
\omega^2
sc,\; \frac{1}{r}(\frac{1}{\alpha^2}+\omega^2s^2),\; 0)
\end{eqnarray}
where
\begin{eqnarray}
{\cal S}_1&=&\sqrt{1-\frac{r^2}{\alpha^2}-\omega^2r^2s^2}\\
{\cal S}_2&=&(1-\frac{r^2}{\alpha^2})(\frac{1}{\alpha^2}+\omega^2s^2)^2
+\omega^4s^2c^2\,.
\end{eqnarray}

It is easy to see that there is no analog of the Keplerian orbits. This is
related to the fact that the `potential $g_{00}$'
is proportional to a positive power of $r$ rather than a negative power as in
the Schwarzschild case. An arbitrary $\omega$ leads to precession
analogous to Thomas precession in flat space but more complicated due to
the curvature of the spatial sections.

\section{Stationary Cylindrically Symmetric Spacetimes}
In this section we extend the treatment of the previous section to
spacetimes which in addition to the Killing vector {\boldmath $\xi$} and
{\boldmath $\eta$} have
yet another Killing vector {\boldmath $\mu$} representing translation
invariance in
the $z$ direction. A well-known example is the G\"{o}del solution
as well as metrics representing solutions with cylindrical symmetry. As
discussed in the beginning all our earlier results obtain in this instance
and hence we write down without proof the main expressions.

We start with the  standard form of the line element  in
this case as given by:
\begin{equation}
ds^2\,=\,g_{00}
dt^2\,+\,2g_{03}\,dtd\phi\,+\,g_{33}\,d\phi^2\,+\,2g_{02}\,dtdz\,+\,g_{22}
\,dz^2\,+\,g_{11}\,d\rho^2\label{eq:51}
\end{equation}
where $g_{ab}$ are functions of $\rho$ only, since we are in coordinates
adapted to the Killing vectors {\boldmath $\xi,\,\eta $} and {\boldmath $\mu$}.
In this case we have,
\begin{eqnarray}
g\,&\equiv&\,\det(g_{ab})\,=\,g_{11}\Delta_{23}\label{eq:52}\\
{\mbox{\rm where} }\;\,
\Delta_{23}\,&\equiv&\,g_{00}g_{33}g_{22}\,-\,g_{22}g^{2}_{03}\,-\,
g_{33}g^{2}_{02}\label{eq:53}
\end{eqnarray}
Further,
\begin{eqnarray}
\begin{array}{ccc}
g^{ab}&=&\frac{{\textstyle 1}}{{\textstyle\Delta_{23}}}\left(
\begin{array}{cccc}
g_{22}\,g_{33}&-g_{03}\,g_{22}&-g_{02}\,g_{33}&0\\
-g_{03}\,g_{22}&\Delta_2&g_{02}g_{03}&0\\
-g_{02}\,g_{33}&g_{02}g_{03}&\Delta_3&0\\
0&0&0&\Delta_{23}/g_{11} \end{array} \right)\label{eq:54}
\end{array}
\end{eqnarray}
\begin{equation}
\Delta_3\,\equiv\,g_{00}\,g_{33}\,-\,g^{2}_{03}\;\;;\;\;
\Delta_2\,\equiv\,g_{00}\,g_{22}\,-\,g^{2}_{02}\label{eq:55}
\end{equation}

Proceeding as before we first compute $\kappa,\,\tau_1$ and $\tau_2$ for an
observer whose world
line is {\boldmath $\xi$}. We obtain
\begin{eqnarray}
\kappa^2\,&=&\,-\,\frac{g^{11}}{4}\,\frac{g^{2}_{00,1}}{g^{2}_{00}}\label
{eq:56}\\
\tau^{2}_{1}\,&=&\,\frac{-g^{11}}{4\Delta_{23}g^{2}_{00}}\left[
g_{00}(g_{02}g_{03,1}-g_{03}g_{02,1})^2\right.-\nonumber\\
&- &\left.g_{22}(g_{03}g_{00,1}-g_{00}g_{03,1})^2
-g_{33}(g_{02}g_{00,1}-g_{00}g_{02,1})^2\right]\label{eq:57}\\
\tau^{2}_{2}\,&=&\,0\label{eq:58}
\end{eqnarray}
For completeness we also compute the  Frenet-Serret bases for these metrics.
It is given by,
\begin{eqnarray}
e_{(0)}^a\,&=&\,\frac{1}{\,\sqrt{g_{00}}}\,(1,\,0,\,0,\,0\,)\nonumber\\
e_{(1)}^a\,&=&\,(\,0,\,\sqrt{-g^{11}},\,0,\,0\,)\nonumber\\
e_{(2)}^a\,&=&\,-\frac{\sqrt{-g^{11}}}{2\sqrt{g_{00}}\Delta_{23}\tau_1}\,
(a_2,\,0,\,b_2,\,c_2)\nonumber\\
e_{(3)}^a\,&=&\,\frac{\sqrt{-g^{11}}}{2g_{00}\sqrt{\Delta_{23}}\tau_1}\,
(a_3,\,0,\,b_3,\,c_3)\label{eq:59}
\end{eqnarray}
where
\begin{eqnarray}
a_2\,&=&\,\frac{1}{g_{00}}\left[g_{22}g_{03}(g_{03}g_{00,1}-g_{00}g_{03,1})+
g_{33}g_{02}(g_{02}g_{00,1}-g_{00}g_{02,1})\right]\nonumber\\
     & &\\
b_2\,&=&\,g_{33}(g_{00}g_{02,1}-g_{02}g_{00,1})+g_{03}(g_{02}g_{03,1}-g_{03}
g_{02,1})\\
c_2\,&=&\,g_{22}(g_{00}g_{03,1}-g_{03}g_{00,1})+g_{02}(g_{03}g_{02,1}-g_{02}
g_{03,1})\\
a_3\,&=&\,g_{03}g_{02,1}-g_{02}g_{03,1}\\
b_3\,&=&\,g_{00}g_{03,1}-g_{03}g_{00,1}\\
c_3\,&=&\,g_{02}g_{00,1}-g_{00}g_{02,1}
\end{eqnarray}

It should be noted that since $\tau_2\,=\,0\;,\;  {\bf e}_{(3)}$ cannot be
obtained
by the usual Frenet-Serret process of differentiation but in this case has
been obtained just by orthonormality with the ${\bf e}_{(i)}\;(i=0,1,2)$.
Adapting the procedure of section (2) to the quasi-Killing congruence
\begin{eqnarray}
\zeta^a\,&\equiv&\,\xi^a\,+\,\omega\eta^a\,+\,v\mu^a\,\equiv\,e^{-\psi}\,
e^{a}_0\\
{\mbox{\rm where} }\;\; \raisebox{-.25cm}{$\stackrel{\textstyle{\cal
L}}{\scriptstyle \zeta}$}\;\omega \;\;&=&\;\;
\raisebox{-.25cm}{$\stackrel{\textstyle{\cal L}}{\scriptstyle
\zeta}$}\;v\;\;=\;\;0
\end{eqnarray}
we obtain,
\begin{eqnarray}
\dot{e}^{a}_{(m)}\,&=&\,F^{a}_{\;\;b}\,e^{b}_{(m)}\\
{\mbox{\rm where} }\;\;
F_{ab}\,&\equiv&\,e^\psi(\xi_{a;b}\,+\,\omega\eta_{a;b}\,+\,v\mu_{a;b})\\
{\rm  and }\;\; \Omega^a\,&=&\,\omega^a\,+\,\tilde{H}^{ab}\,e_{(0)b}\\
{\mbox{\rm where} }\;\; \omega^a\,&=&\,\tilde{F}^{ab}\,e_{(0)b}\;\; {\rm
and}\;\;H_{dc}\,\equiv\,e^\psi\,[\omega_{[,d}\eta_{c]}\,+\,v_{[,d}\mu_{c]}
\end{eqnarray}

The general quasi-Killing trajectories along {\boldmath $\zeta$} represent
helical
orbits. Nevertheless, the computation of $\kappa,\,\tau_1,\, \tau_2$ for
$\zeta^a$ involves a
similar trick as before. Under the coordinate transformation
\begin{eqnarray}
t\,&=&\,t^\prime\\
\phi\,&=&\,\phi^\prime\,+\,\omega t^\prime\\
z\,&=&\,z^\prime\,+\,vt^\prime
\end{eqnarray}
the metric transforms to,
\begin{eqnarray}
g_{0'0'}\,&=&\,g_{00}\,+\,2\omega
g_{03}\,+\,\omega^2g_{33}\,+\,2vg_{02}\,+\,v^2g_{22}\,\equiv\,{\cal D}\\
g_{0'3'}\,&=&\,g_{03}\,+\,\omega g_{33}\,\equiv\,{\cal B}\\
g_{0'2'}\,&=&\,g_{02}\,+\,vg_{22}\,\equiv\,{\cal E}\\
g_{3'3'}\,&=&\,g_{33}\;\;\;;\;\;\;g_{2'2'}\,=\,g_{22}
\end{eqnarray}
where $g_{a'b'}$ are independent of $t'\,, \phi '\,$  and $z'$. The Killing
vector $\mbox{\boldmath $\zeta $}\,=\,(1, 0, 0, 0)$ corresponds in the old
coordinates to
$\mbox{\boldmath $ \xi $}\,+\,\omega\mbox{\boldmath $ \eta $}\,+\,
v\mbox{\boldmath $ \mu $}$ and consequently we can use
eqs.(\ref{eq:56}-\ref{eq:58}) to
evaluate $\kappa,\,\tau_1,\,\tau_2$ by using $g_{a'b'}$ instead of $g_{ab}$
in the equations. This gives,
\begin{eqnarray}
\kappa^2\,&=&\,-\,\frac{g^{11}\,{\cal D}_{(1)}^2}{4\,{\cal D}^2}\\
\tau^{2}_{1}\,&=&\,\frac{-g^{11}}{4\Delta_{23}{\cal D}^2}\left[{\cal D}({\cal
EB}_{(1)}-{\cal BE}_{(1)})^2\right.\nonumber\\
&- &\left.g_{22}({\cal BD}_{(1)}-{\cal DB}_{(1)})^2
-g_{33}({\cal ED}_{(1)}-{\cal DE}_{(1)})^2\right]\\
\tau^{2}_{2}\,&=&\,0\\
\mbox{{\rm where}}\;\;
{\cal D}_{(1)}\,&\equiv &\,g_{00,1}\,+\,2\omega
g_{03,1}\,+\,\omega^2g_{33,1}\,+\,2vg_{02,1}\,+\,v^2g_{22,1}\\
{\cal E}_{(1)}\,&\equiv&\,g_{02,1}\,+\,vg_{22,1}
\end{eqnarray}
\begin{eqnarray}
e^{a}_{(0)}&=&\frac{1}{\sqrt{{\cal D}}}\,(1,\,0,\,v,\,\omega)\nonumber\\
e^{a}_{(1)}&=&(0,\,\sqrt{-g^{11}},\,0,\,0)\nonumber\\
e^{a}_{(2)}&=&-\frac{\sqrt{-g^{11}}}{2\sqrt{{\cal
D}}\Delta_{23}\tau_1}\,(a'_{2},\,0,\,b'_{2}+va'_{2},\,c'_{2}+\omega
a'_{2})\nonumber\\
e^{a}_{(3)}&=&\frac{\sqrt{-g^{11}}}{2{\cal
D}\sqrt{\Delta_{23}}\tau_1}\,(a'_{3},
\,0,\,b'_{3}+va'_{3},\,c'_{3}+\omega a'_{3})
\end{eqnarray}
where $a'_{i}$ and $b'_{i}$ $(i=2,3)$ refer to $a_i$ and $b_i$ with
$g_{ab}$ replaced by the corresponding $g_{a'b'}$.

The geodesics are determined by $\kappa=0$ {\em i.e.~} ${\cal D}_{(1)}\,=\,0$
and in this case $\tau_1^2$ simplifies to:
\begin{eqnarray}
\tau_1^2\,&=&\,\frac{g_{11}}{4\Delta_{23}{\cal D}}\left[
(g^{2}_{02}-g_{22}{\cal A}){\cal B}^{2}_{(1)}+\right.\nonumber\\
&+&\left.(g^{2}_{03}-g_{33}{\cal F}){\cal E}^{2}_{(1)}
-2{\cal EB}\,\,{\cal B}_{(1)}{\cal E}_{(1)}\right]\\
{\rm where}\;\; {\cal F}\,&\equiv&\,g_{00}\,+\,2vg_{02}\,+\,v^2g_{22}
\end{eqnarray}

A further simplification obtains if the spacetimes under consideration satisfy
\begin{equation}
g_{02,\rho}\,=\,g_{22,\rho}\,=\,0
\end{equation}
In this case the Keplerian orbits for {\boldmath $\zeta$} are determined by
${\cal A}_{(1)}
\,=\,0$ as for the {\boldmath $\chi$} congruence and $\tau_1^2$ reduces to:
\begin{equation}
\tau_1^2\,=\,-\frac{g^{11}{\cal B}^{2}_{(1)}(g^{2}_{02}-g_{22}{\cal
A})}{4\Delta_{23}{\cal D}}
\end{equation}
The further restriction $g_{02}\,=\,0$ finally leads to a form useful for
the G\"{o}del case:
\begin{equation}
\tau_1^2\,=\,\frac{g^{11}{\cal B}_{(1)}^{2}{\cal A}}{4\Delta_3({\cal A}\,
+\,v^2 g_{22})}
\end{equation}
The general expressions above may be simplified in three particular cases
(i)~ $g_{02}=0$ (ii) $g_{02,\rho}=g_{22,\rho}=0$ (iii)
$g_{02}=0\,\,\,g_{22,\rho}=0$. The G\"{o}del Universe belongs to category (iii)
while the cylindrical vacuum metrics of \cite{vw77} belong to class (i).

We next apply these general considerations to the G\"{o}del case and finally,
to the general
cylindrically symmetric vacuum metrics.
\subsection{ G\"{o}del Universe}
The G\"{o}del Universe is described by the line element
\begin{equation}
ds^2\,=\,4R^2\,[dt^2\,+\,2\sqrt{2}\,S^2d\phi
dt\,-\,(S^2-S^4)\,d\phi^2-dr^2-dz^2]
\end{equation}
where $S\,\equiv\,\sinh\,r\;\;;\;\;C\,\equiv\,\cosh\,r$.
\subsubsection{General Spiralling \protect{\boldmath $\zeta$}- Trajectories}
Adapting the formulas of the previous section for observers moving along
{\boldmath $\zeta\;\;$} we have,
\begin{eqnarray}
\kappa^2\,&=&\,\frac{\omega^2S^2C^2\left[2\sqrt{2}\,-\,\omega(1-2S^2)\right]^2}
{4R^2{\cal G}_1^2}
\label{eq:goz1}\\
\tau^{2}_{1}\,&=&\,\frac{1}{4R^2{\cal G}_1^2}\{{\cal G}_2^2\,+\,v^2({\cal
G}_3^2\,+\,{\cal G}_4^2{\cal G}_1)\}\\
\tau^{2}_{2}\,&=&\,0\\
{\rm where}\;\;{\cal G}_1\,&\equiv&\,1\,+\,2\sqrt{2}\omega
S^2\,-\,\omega^2S^2(1-S^2)\,
-\,v^2\\
{\cal
G}_2\,&\equiv&\,\omega(1-2S^2)\,-\,\sqrt{2}(1+\omega^2S^4)\,+\,v^2\{\sqrt{2}-
\omega
(1-2S^2)\}\nonumber\\
&&\\
{\cal G}_3\,&\equiv&\,\omega S\sqrt{1-S^2}\{2\sqrt{2}-\omega(1-2S^2)\}\\
{\cal G}_4\,&\equiv&\,\sqrt{2}-\omega(1-2S^2)\label{eq:goz2}
\end{eqnarray}
\subsubsection{Geodesics}
Along a geodesic $\kappa\,=\,0$ yielding for the `Keplerian' frequency
\begin{eqnarray}
\omega\,=\,\frac{2\sqrt{2}}{1-2S^2}
\end{eqnarray}
the same as for the {\boldmath $\xi$}-orbits \cite{rp90}.
However, the gyroscopic precession frequency contains the signature
of the $z-$motion and is given by
\begin{eqnarray}
\tau^{2}_{1}\,=\,\frac{1-4S^2C^2}{2R^2[1-4S^2C^2\,-\,v^2(1-2S^2)^2]}\,,
\end{eqnarray}
\subsubsection{The \protect{\boldmath $\chi$}- Motion}
The $v=0$ limit of the eqs.(\ref{eq:goz1}-\ref{eq:goz2}) leads us
to the motion along {\boldmath $\chi -$}lines and we obtain:
\begin{eqnarray}
\kappa^2\,&=&\,\frac{\omega^2S^2C^2}{4R^2}\,\left[\frac{2\sqrt{2}\,-\,
\omega(1-2S^2)}{1+2\sqrt{2}
\omega S^2\,-\,\omega^2S^2(1-S^2)}\right]^2\\
\tau^{2}_{1}\,&=&\,\frac{(\sqrt{2}\,-\,\omega(1-2S^2)\,+\,\sqrt{2}\,
\omega^2S^4)^2}{4R^2[1+2\sqrt{2}\omega
S^2\,-\,\omega^2S^2(1-S^2)]^2}
\end{eqnarray}

The precession frequency for motion along circular geodesics takes the
simple form,
\begin{eqnarray}
\tau^{2}_{1}\,=\,\frac{1}{2R^2}
\end{eqnarray}
yielding for the precession,
\begin{eqnarray}
\Delta\phi\,=\,-\pi[(1-\sinh^2 2r)^{1/2}\,-\,2]
\end{eqnarray}
in agreement with earlier results \cite{rp90}.

Finally in the case of stationary observers({\boldmath $\xi$}-lines) $\kappa=0$
{\em i.e}
the $t$-lines are geodesics. In this case we obtain
\begin{equation}
\tau_1^2\;=\;\frac{1}{2R^2}.
\end{equation}
Thus the Frenet-Serret frame of the stationary observers precesses
relative to the gyroscopes and reveals the rotation intrinsic to the G\"{o}del
universe. Following the procedure outlined in sec.4.1.3 the precession due
to dragging is
\begin{equation}
\Delta\phi_{\tt (drag)}\,=\,\pi(\,1\,-\,2\sinh^2 r)
\end{equation}
in agreement with \cite{rp90}.
\subsubsection{GHOST}
We conclude by a consideration of precession along GHOST trajectories.
The angular velocity
of these observers correspond to
\begin{eqnarray}
\omega\,=\,-\,\frac{g_{03}}{g_{33}}\,=\,\frac{\sqrt{2}}{1-S^2}
\end{eqnarray}
leading to
\begin{eqnarray}
\kappa^2\,&=&\,\frac{S^2}{R^2C^2(1-S^2)^2}\\
{\rm and}\;\;\tau_{1}^{2}\,&=&\,\frac{S^4}{2R^2(1-S^2)^2}
\end{eqnarray}
\subsection{Stationary cylindrically symmetric vacuum spacetimes}
The stationary cylindrically symmetric vacuum spacetimes have been given
in an elegant compact form by Vishveshwara and Winicour \cite{vw77} as
\begin{eqnarray}
ds^2\,&=&\,e^{2\varphi}\,(d\tau^2\,+\,d\sigma^2)\,+\,\lambda_{00}dt^2\,+\,
2\lambda_{03}dtd\phi+\,\lambda_{33}d\phi^2\\
\mbox{{\rm where} }\;\,
\lambda_\alpha\,&=&\,A_\alpha\,\tau^{1+b}\,+\,B_\alpha\,\tau^{1-b}\;\;;\;\;
\alpha\,=\,00,\,03,\,33\\
e^{2\varphi}\,&=&\,c\,\tau^{b^{2}-1}\;;\;\;\tau\,=\,\sqrt{2}\rho\;;\;\;\sigma\,
=\,\sqrt{2}z
\end{eqnarray}
The coefficients $A_{\alpha}$ and $\beta_{\alpha}$ satisfy the following
algebraic relations:
\begin{eqnarray}
A_{00}\,A_{33}\,-\,A^{2}_{03}\,=\,B_{00}\,B_{33}\,-\,B^{2}_{03}\,&=&\,0\,\\
A_{00}\,B_{33}\,+\,A_{33}\,B_{00}\,-\,2A_{03}\,B_{03}\,&=&\,-\frac{1}{2}
\end{eqnarray}
The mass per unit length $m$ and angular momentum per unit length $j$ are
given by
\begin{eqnarray}
m\,&=&\,\frac{1}{4}\,+\,\frac{1}{2}\,b(A_{33}B_{00}\,-\,A_{00}B_{33})\\
j\,&=&\,\frac{1}{2}\,b(A_{03}B_{33}\,-\,A_{33}B_{03})
\end{eqnarray}
\subsubsection{The \protect{\boldmath $\zeta$}-Trajectories}
For completeness we write down the Frenet-Serret invariants for the line
element listed above. They turn out to be,
\begin{eqnarray}
\kappa^2\,&=&\,-\frac{1}{4c\tau^{b^2+1}}
\left[\frac{(1+b)\tau^bA_{0'0'}+(1-b)\tau^{-b}B_{0'0'}+(b^2-1)v^2c\tau^{b^2-2}}
{A_{0'0'}\tau^b+B_{0'0'}\tau^{-b}+v^2c\tau^{b^2-2}}\right]^2\nonumber\\
          & &\\
\tau_{1}^2\,&=&\,\frac{1}{\tau^3{\cal W}^2}\left[\frac{-2{\cal W}_1^2}
{c\tau^{b^2-2}}+v^2({\cal W}{\cal W}_3^2-{\cal W}_2^2)\right]
\end{eqnarray}
where
\begin{eqnarray}
A_{0'0'}\,&\equiv&\,A_{00}+2\omega A_{03}+\omega^2A_{33}\\
B_{0'0'}\,&\equiv&\,B_{00}+2\omega B_{03}+\omega^2B_{33}\\
A_{0'3'}\,&\equiv&\,A_{03}+\omega A_{03}\;;\;\;B_{0'3'}\,\equiv\,B_{03}+\omega
B_{03}\\
{\cal W}\,&\equiv&\,A_{0'0'}\tau^b+B_{0'0'}\tau^{-b}+2v^2c\tau^{b^2-2}\\
{\cal
W}_1\,&\equiv&\,b\left[A_{0'0'}B_{0'3'}-A_{0'3'}B_{0'0'}\right]-\nonumber\\
&&-v^2c\tau^{b^2-2}\{(1+b)(2-b)A_{0'3'}\tau^b+(1-b)(2+b)B_{0'3'}\tau^{-b}\}
\nonumber
\\
&&\\
{\cal W}_2\,&\equiv&\,\sqrt{A_{33}\tau^b+B_{33}\tau_{-b}}\left[
(1+b)(2-b)A_{0'0'}\tau^{b}+\right.\nonumber\\
&&\left.+(1-b)(2+b)B_{0'0'}\tau^{-b}\right]\\
{\cal W}_3\,&\equiv&\,(1+b)(2-b)A_{0'3'}\tau^b+(1-b)(2+b)B_{0'3'}\tau^{-b}
\end{eqnarray}
\subsubsection{Observers with arbitrary constant angular velocity along
\protect{\boldmath ${\bf\chi}$}}
In this case we obtain
\begin{eqnarray}
\kappa^2\,&=&\,-\frac{1}{4c\tau^{b^{2}+1}}
\left[\frac{(1+b)A_{0'0'}\tau^b\,+\,(1-b)B_{0'0'}\tau^{-b}}{A_{0'0'}\tau^b\,+
\,B_{0'0'}\tau^{-b}}\right]^2\\
\tau^{2}_{1}\,&=&\,-\frac{2b^2}{c\tau^{b^{2}+1}}
\left[\frac{(A_{0'0'}B_{0'3'}\,-\,A_{0'3'}B_{0'0'})}{A_{0'0'}\tau^b\,+
\,B_{0'0'}
\tau^{-b}}\right]^2\,.
\end{eqnarray}
\noindent
Note
\begin{eqnarray}
b(A_{0'0'}B_{0'3'}-A_{0'3'}B_{0'0'})&=&b\left[(A_{00}B_{03}-A_{03}B_{00}+
\right.\nonumber\\
&+&\left.\omega(A_{00}B_{33}-A_{33}B_{00})
+\omega^2(A_{03}B_{33}-A_{33}B_{03})\right]\nonumber\\
&=&\frac{1}{2}[2b(A_{00}B_{03}-A_{03}B_{00})-\omega(4m-1)+4j\omega^2]
\nonumber\\
&&
\end{eqnarray}
\subsubsection{Keplerian geodesics}
These are determined by $\kappa\,=\,0$ yielding
\begin{eqnarray}
\omega\,&=&\,\frac{[-(1+b)
A_{03}\tau^b\,+\,(1-b)B_{03}\tau^{-b}]\,\pm\,(\frac{1-b^2}{2})^{\frac{1}{2}}}
{(1+b)A_{33}\tau^b\,+\,(1-b)B_{33}\tau^{-b}}
\end{eqnarray}
Note that real roots are possible only for $b^2 < 1$ which is consistent
with the fact that the potential $g_{00}$ is a function of negative powers
of $\tau$ only for these values. The precession is obtained to be
\begin{eqnarray}
\tau^{2}_{1}\,=\,-\frac{(1-b^2)}{4c\,\tau^{1+b^{2}}}
\end{eqnarray}
{\subsubsection{GHOST}
Finally we look at these special trajectories for the cylindrically
symmetric vacuum metrics. In this instance
\begin{eqnarray}
\omega\,=\,-\,\frac{g_{03}}{g_{33}}\,=\,-\,\frac{A_{03}\,\tau^b\,+
\,B_{03}\,\tau^{-b}}{A_{33}\,\tau^b\,+\,B_{33}\,\tau^{-b}}
\end{eqnarray}
and the acceleration becomes
\begin{eqnarray}
\kappa^2=-\frac{1}{4c\tau^{b^{2}+1}}\left[\frac{(1-b)A_{33}\tau^b+(1+b)B_{33}
\tau^{-b}}
{A_{33}\tau^b\,+\,B_{33}\tau^{-b}}\right]^2
\end{eqnarray}
The precession takes the form
\begin{eqnarray}
\tau^{2}_{1}\,=\,-\,\frac{8j^2}{c\tau^{b^{2}+1}\,(A_{33}\tau^b\,+\,
B_{33}\tau^{-b})^2}
\end{eqnarray}
The gyroscopic precession given by $\tau_1$ is proportional in this case
to the specific angular momentum $j$. From our previous discussion we know
that the connecting vector between two adjacent trajectories of the
`irrotational' congruence does not precess with respect to the
Fermi-Walker transported gyroscope. In general, the Frenet-Serret triad
does precess with respect to the gyroscope or equivalently in this case
with respect to the connecting vector. However, if $\tau_1\,=\,0$ ($\tau_2$
is identically zero) then the Frenet-Serret triad is also non precessing
with respect to the gyroscope or the connecting vector. This happens when
the angular momentum of the source $j\,=\,0$. Thus, the observer can
decide if the source is rotating or not by checking whether his
Frenet-Serret triad - which is Lie transported - precesses or not with
respect to the gyroscope or the connecting vector. Gyroscopic precession along
trajectories of the irrotational congruence reveal directly the
rotation of the central source.
\subsubsection{The stationary observers}
For completeness we write down the parameters when $\omega=0$, {\it i.e.,}
the {\boldmath $\xi$}-lines. We have
\begin{eqnarray}
\kappa^2&=&-\frac{1}{4c\tau^{b^2+1}}\left[\frac{(1+b)A_{00}\tau^b+
(1-b)B_{00}\tau^{-b}}{A_{00}\tau^b+B_{00}\tau^{-b}}\right]^2\\
\tau^{2}_{1}&=-&\frac{2b^2}{c\tau^{b^{2}+1}}\left[\frac{A_{00}B_{03}-
A_{03}B_{00}}{A_{00}\tau^b+B_{00}\tau^{-b}}\right]^2
\label{eq:fin}
\end{eqnarray}
Eq.(\ref{eq:fin}) describes the rotation of the stationary gyroscope due
to dragging. Once again this illustrates the very general effect of spacetime
rotation on local experiments.
\section{Conclusion}
As has been mentioned earlier, gyroscopic precession is a phenomenon that
has been extensively studied both in flat and curved spacetimes by
different methods. The orbits of the gyroscopes in these instances are
given by combinations of Killing directions admitted by the spacetimes
under consideration. It is found that in these circumstances, the
invariant geometrical description of the Frenet-Serret formalism provides
a covariant and elegant framework for the study
of precession. By extending the formalism applied to Killing trajectories
to quasi-Killing trajectories, a large number of cases can be studied in
a unified manner. Furthermore, this treatment makes it
possible to relate the precession of a gyroscope to the vorticity of a
congruence when the gyroscope is transported along a given member of that
congruence. An important example is the irrotational congruence admitted
by stationary, axisymmetric spacetimes like the Kerr. It is worth
pointing out, however, that gyroscopic precession is directly determined
by the Frenet-Serret rotation in general. Another aspect of our treatment
is the unified description of precession applicable to a whole family of
spacetimes. Specifically, precession for orbits with  arbitrary constant
angular
speed has been worked out for the Kerr metric. Starting from this,
particular examples have been worked out for the entire
Kerr-Schwarzschild-Minkowski
spacetimes. Expressions presented are general, exact
and not confined to the equatorial plane. In deriving these results
rotating coordinate systems have been used to generate circular orbits
from static trajectories. Also, additional interesting cases such as
G\"{o}del, De Sitter universe and general vacuum cylindrical spacetimes have
been investigated. It would be interesting to explore in detail the
implications of the general results obtained here and their possible
astrophysical applications.


\begin{thebibliography}{96}
\bibitem {app} Along a general congruence the evolution of the
connecting vector is governed by the quantities vorticity,
shear and expansion. In the case of a Killing
congruence only the vorticity is non-zero. None
of the three quantities is zero in the case of an arbitrary
quasi-Killing congruence.
However,  axisymmetric stationary spacetimes, such as the
Kerr spacetime, admit an irrotational quasi-Killing congruence for which
the vorticity is zero. Expansion would be zero if the coefficient $\omega_A$
is Lie derived by the Killing vector {\boldmath $\eta_A$} associated with it.
\end{thebibliography}
\end{document}